\documentstyle[psfig,aps,prc,preprint,floats]{revtex}   % Use for submission.
\tighten                                  % Single line spacing

\begin{document}
\preprint{ECT$^*$-97-002, CU-TP-814/97, 
          DPNU-97-11, LUNFD6/(NFFL-7137)1997}
\draft
\title{  Quantum Statistical Correlations and \\
         Single Particle Distributions for    \\
         Slowly Expanding Systems with
         Temperature Profile                     }
\author{      J.\ Helgesson$^{1}$,
              T.\ Cs\"org\H o$^{2,3}$,
              M.\ Asakawa$^{4}$,
         and  B.\ L\"orstad$^{5}$
         \thanks{ helgesso@ect.unitn.it,
                  csorgo@rmki.kfki.hu,
                  yuki@nuc-th.phys.nagoya-u.ac.jp,
                  bengt@quark.lu.se }                }

\address{ $^1$ ECT$^*$, European Centre for Studies in
                Theoretical Nuclear Physics \\
                and Related Areas, Villa Tambosi,\\
		Strada delle Tarabelle 286,
	        I-38050 Villazzano (Trento), Italy\\
           $^2$ MTA KFKI RMKI,  H--1525 Budapest 114,
                P.O. Box 49, Hungary, \\
           $^3$ Department of Physics, Columbia University, \\
                538 W 120th Street, New York, NY  10027, USA, \\
           $^4$ Department of Physics, School of Science,
                Nagoya  University,  \\
	        Nagoya, 464-01, Japan \\
           $^5$ Department of Physics, University of Lund, \\
                Box 118, S - 221 00 Lund, Sweden \\              }
\date{February 11, 1997}

\maketitle
\begin{abstract}
Competition among particle evaporation,
temperature gradient and flow
is investigated in a phenomenological manner,
based  on a simultaneous analysis
of quantum statistical correlations
and momentum distributions
for a non-relativistic, spherically symmetric,
three-dimensionally expanding, finite source.
The parameters of the  model emission function
are constrained by fits
to neutron and proton momentum distributions
and correlation functions
in intermediate energy heavy-ion collisions.
The temperature gradient is related
to the momentum dependence of the radius parameters
of the two-particle correlation function,
as well as to the  momentum-dependent temperature parameter
of the single particle spectrum,
while a long duration of particle evaporation
is found to be responsible for the low relative momentum behavior
of the two-particle correlations.
\end{abstract}


% ---------------------------------------------------------------

% ---------------------------------------------------------------
\section{Introduction}
\label{sec_introd}
% ---------------------------------------------------------------

It has been shown recently that for non-relativistic,
three-dimensionally expanding systems
the quantum statistical correlations
measure only part of the particle source
and that the effective temperature
of the momentum distribution
is obtained as a combination
of the freeze-out temperature 
and of the ``geometrical temperature'',
a term due to expansion 
and to the finite geometrical sizes \cite{nr}.
The expansion makes the effective radius parameters
of the two-particle correlation functions
smaller than the geometrical size of the source
even for non-relativistic expanding systems \cite{nr}.

Here we extend our study
of non-relativistic expanding particle sources
to the case of a temperature gradient inside the source.
Such  an extension of the model of Ref.\ \cite{nr}
is motivated by data on momentum distribution
of protons in heavy ion reactions
in the non-relativistic energy domain,
which  show deviations
from the purely exponential spectrum \cite{chitwood,bo},
although the neutron spectrum is well approximated
by a Boltzmann distribution.

The purpose of our paper is to investigate the features
of non-relativistic heavy ion collisions
at a phenomenological level,
parameterizing the emission function
assuming local thermal equilibrium,
without attempting to describe the microscopic mechanisms
which lead to such a particle emission pattern.
However, we include phenomenologically general features like
evaporation, cooling, temperature gradient and flow
as well as strong and Coulomb final state interactions
for protons and neutrons,
and test the resulting model on data at $E = 30 $ MeV/nucleon.
Analytic approximations are also derived
in order to gain improved insight
into the influences of the competing effects.
These analytic results are not directly applicable
to nucleon correlations.
However, they may be applicable
to study pion correlation functions and spectra
in such reactions where a spherical non-relativistic source
may be assumed.

At high energies a relativistic analogue model \cite{3d}
has been
{successful in describing}
the available data \cite{3ds}.
However there are  important qualitative differences
between relativistic heavy ion collisions at CERN SPS
and those at non-relativistic energies from the point of view
of particle sources.
Low and intermediate energy reactions may create
a very long-lived, evaporative source,
with characteristic lifetimes of a few 100 fm/c,
in contrast to the relatively short-lived systems
of lifetimes not larger than a few 10 fm/c at CERN SPS.
During such long evaporation times,
cooling of the source is unavoidable
and has to be included into the model,
while the freeze-out at a constant temperature
is a much better description at higher energies.
Furthermore, in the non-relativistic heavy ion collisions
mostly protons and neutrons are emitted
and they have much stronger final state interactions
than the pions dominating the final state
at ultra-relativistic energies
(for recent reviews on nucleon interferometry,
 see for example Refs.\ \cite{boal,bauer,ardouin}).

We would like to emphasize that it is not our purpose
to re-invent a detailed microscopical description
of non-relativistic nuclear collisions.
We are trying to develop a framework
to describe the particle emission patterns,
i.e., a simultaneous description
of the invariant momentum distribution
and the quantum statistical correlation function,
from intermediate energies
up to the highest energies achievable.
In this paper we concentrate on the low part
of the intermediate energy range.
For example,
sophisticated microscopical transport descriptions \cite{aichelin},
such as the BUU (Boltzmann-Uehling-Uhlenbeck)
and the QMD (Quantum Molecular Dynamics) models,
are well-known and believed to provide a reasonable picture
of proton emission in central heavy ion collisions
from a few tenths up to hundreds of MeV per nucleon.
However, in Ref.\ \cite{upe},
the BUU model predicts too large correlations
and under-predicts the number of protons
emitted with low energies,
for the reaction $^{36}$Ar + $^{45}$Sc
at E = 120 and 160 MeV/nucleon.
This indicates that the simultaneous description
of two-particle correlations and single-particle spectra
is a rather difficult task.
For energies below a few tens of MeV per nucleon,
where long-lived evaporative particle emission
is expected to dominate,
the measured two-proton correlation functions 
were found to be consistent 
with compound-nucleus model predictions \cite{cmpn};
however, a simultaneous analysis
of proton and neutron single particle spectra
and two-particle correlation
has not yet been performed
{to the best of our knowledge}.

The basic model, including a temperature gradient,
is presented in section \ref{sec_modTGrad}.
Section \ref{sec_LoEmTi} contains an extension
of the model to long emission times.
Analytic approximations for momentum distributions
and correlation functions
are derived in section \ref{sec_anaAppr},
and applied in section \ref{sec_C(q)wNoFSI}
to situations where final state interactions can be ignored.
Section \ref{sec_ppAndnn} contains numerical applications
to neutron and proton interferometry,
where final state interactions are important.
Finally, our results are summarized in section \ref{sec_Sum},
while appendix \ref{sec_appA} contains additional material
on analytical approximations for the momentum distribution.

% ---------------------------------------------------------------
\section{The model}
\label{sec_model}
% ---------------------------------------------------------------
The model presented in this section
is based on the work of Ref.\ \cite{nr}.
For clarity we briefly summarize
the formalism along the lines of Ref.\ \cite{nr},
together with a thorough description of the extensions.

%................................................................
\subsection{Emission Function with a Temperature Gradient}
\label{sec_modTGrad}
%................................................................
The emission is characterized
by the emission function or source function $S(x,p)$
which is the probability that a particle
is produced at a given
$x = (t, \bbox{r}\,) = (t,r_x,r_y,r_z)$
point in space-time with the four-momentum
$p = (E, \bbox{p}\,) = (E, p_x, p_y, p_z)$,
where the particle is on mass shell,
$ m^2 = E^2 - \bbox{p}^{\, 2} $.
The quantum-mechanical analogy
to the classical emission function
is the time-derivative
of the non-relativistic Wigner-function \cite{pratt_csorgo}
which is analogous 
to the covariant Wigner-function of Ref.\ \cite{uli_w}.
In this work the time-derivative of the Wigner-function
will be approximated by classical emission functions.

In terms of the emission function
both the invariant momentum
{distribution}
(IMD) and the quantum statistical correlation function (QSCF)
are prescribed.
The Fourier-transformed emission function
is introduced as an auxiliary quantity
\begin{eqnarray}
   \tilde S(\Delta k; K )
=
   \int d^4 x \, S(x;K) \, \exp( \, i \Delta k \cdot x \,)\ ,
\end{eqnarray}
where
\begin{eqnarray}
                   \Delta k  =  p_1 - p_2,
& \qquad\qquad &
                   K = \frac{ p_1 + p_2 }{ 2 }
\end{eqnarray}
and
$\Delta k \cdot x $
stands for the four-product of the four-vectors.
Then the momentum distribution
of number of the emitted particles,
$N_1(\bbox{p})$
is given by
\begin{eqnarray}
       N_1(\bbox{p})
& = &
       \frac{ d^3 n }{ dp_x \, dp_y \, dp_z }
  =
       \tilde S(\Delta k = 0, K = p),
\end{eqnarray}
which is normalized to the mean multiplicity as
\begin{eqnarray}
       \int d^3 p \, N_1(\bbox{p})
& = &
       \langle n \rangle.
\end{eqnarray}
Note that this normalization condition
is different from the one used in Ref.\ \cite{nr}
where the momentum distribution has been normalized to unity.
We also assume that the non-relativistic measure
of the momentum space should be applied,
due to the non-relativistic nature of the considered problem.

In the plane-wave approximation
(i.e.\ neglecting final state interactions),
the Bose-Einstein or Fermi-Dirac correlation functions
are prescribed in terms of our auxiliary function as
\begin{eqnarray}
            C(K,\Delta k)
& \simeq &
            1 \pm \frac{ \mid \tilde S(\Delta k , K) \mid^2 }
                       { \mid \tilde S(0,K) \mid^2 },
\end{eqnarray}
where the $+$ sign stands for bosons
and the $-$ sign for fermions.
This approximation involves an off-shell continuation of the
on-shell emission functions, the significance of which
was discussed first in Refs.\ \cite{pratt_csorgo,zajc}.

For central heavy ion collisions at intermediate energies
the target and the projectile
form a collective state which can be described
as a non-relativistically expanding fluid
within the framework of hydrodynamical models.
Due to the expansion, the fluid cools
and we assume that it then  
undergoes a certain disintegration process.
In case the information about the initial directions is lost,
the final freeze-out stage 
becomes approximately spherically symmetric.

We assume  along the lines of Ref.\ \cite{nr}
that the emission function is characterized
by a distribution of production points $I(\bbox{r}\,)$
and by a distribution of the particle emission times, $H(t)$.
The correlations between space-time and momentum-space
shall be introduced
by a non-relativistic momentum distribution.
We assume that the expanding system is dilute enough
when the particles are emitted
so that the quantum statistical single-particle distribution
can be well approximated by a Boltzmann distribution,
\begin{eqnarray}
        f(x;p)
& = &
        \frac{ g }{  (2 \pi  )^{3} } \,
        \exp \left( - \frac{ (\bbox{p} - m \bbox{u}(x))^2 }
                           { 2 m T(x) }
             \right) I(x \,),
\\
        I(x \,)
& = &
        \exp\left( \frac{ \mu(x) }{ T(x) } \right).
\end{eqnarray}
Here $g$ is the degeneracy factor,
$\bbox{u}(x) $ is the (non-relativistic) flow velocity,
the freeze-out temperature is denoted by $T$
and $\mu(x) $ is the chemical potential.

Thus the emission function is characterized as
\begin{eqnarray}
S(x;K) & = &  f(x;K) \,  H(t).
\end{eqnarray}
In order to simplify the results
we shall keep only the mean and the width
of the source distributions,
i.e.,  we shall apply the Gaussian approximations
for the distribution functions
of $ t$ and $\bbox{r}$ as follows
\begin{eqnarray}
        I(x\,)
& = &
        \exp\left(  \frac{ \mu_0 }{ T_0 } \right)
        \exp\left( - \frac{ {\bbox{r}\,}^2 }{ 2 R_G^2 } \right),
\\
        H(t)
& = &
        \frac{ 1 }{ (2 \pi \Delta t^2 )^{1/2} } \,
        \exp\left( - \frac{ (t - t_0)^2 }{ 2 \Delta t^2 } \right).
        \label{eq_H(t)_Gauss}
\end{eqnarray}
(The expression in Eq.\ (\ref{eq_H(t)_Gauss})
 is not suitable for very long emission times
 and evaporative processes.
 Such scenarios will be treated and discussed
 in the next section.)

In other words,
we have the following ansatz
for the chemical potential $\mu(x)$:
\begin{eqnarray}
        \frac{ \mu(x) }{ T(x) }
& = &
        \frac{ \mu_0 }{ T_0 } - 
        \frac{ \bbox{r}^{\, 2} }{ 2 R_G^2 },
\end{eqnarray}
which is analogous to the ones used in Refs.\ \cite{nr,uli}.

We assume the following form
of the local temperature distribution:
\begin{eqnarray}
    T(\bbox{r})
=
    \frac{ T_0 }{ 1 + a^2 \, \bbox{r}^2 / 2 t_0^2 }\ .
\label{e:tprof}
\end{eqnarray}
The parameter $a$ controls the gradient
of the local temperature at the last interaction point.
We will in this work only treat
the case when the temperature profile 
is decreasing as a function of $r$,
though the expression (\ref{e:tprof})
contains also increasing temperature profiles
by taking $a$ pure imaginary.
   Note, that the particular choice
   of the temperature profile
   is not influencing the leading order, approximate results:
   namely, we shall apply a saddle-point approximation
   and an expansion of the inverse temperature profile
   around the $\bbox{r} = 0, t = t_0$ point.
   Thus any other temperature profile,
   which leads to similar expansion coefficients,
   shall lead to similar approximate results.

We assume that the freeze-out temperature
at each emission point
is much smaller than the mass of the particles
in this non-relativistic case,
\begin{eqnarray}
        T(\bbox{r},t) & \ll & m.
\end{eqnarray}

We select a velocity of the 3D expanding matter
at space-time point $x$
so that it be spherically symmetric
and describe an expansion in all three directions
with a constant gradient.
Thus the velocity around the mean freeze-out time $t_0$
is assumed to have the form
\begin{eqnarray}
     \bbox{u}(x) & = & b \, \frac{ \bbox{r} }{ t_0 },
\end{eqnarray}
which describes a scaling solution
of the non-relativistic hydrodynamical equations
at the mean freeze-out time $t_0$ for
$\mid \!\bbox{r}\! \mid  \ll t_0$ and $b =1$,
Ref.\ \cite{jozso}.
Here we have introduced the parameter $b$
which controls the amount of flow.
For $b = 0$ we recover the case without flow.
The results given in Ref.\ \cite{nr}
are re-obtained for the case $a = 0$ and $b = 1$.

In Ref.\ \cite{nr} it was implicitly assumed
that the duration of the particle emission is short,
$ \Delta t \ll t_0$,
since the flow field,
the geometrical radius and the freeze-out temperature
were assumed not to change significantly 
during the time interval $\Delta t$,
centered on $t_0$.
The duration of the particle emission, $\Delta t$,
has thus to satisfy
\begin{eqnarray}
        \Delta t \ll \min \left( \frac{ t_0 }{ a },\,\,
                                \frac{ t_0 }{ b } \right)
\end{eqnarray}
in order to warrant the model assumptions.
If we have
$ a, b \ll 1$
then
$\Delta t \approx t_0$
becomes possible since we approach the static fire-ball case.

When the above approximations for the emission function are valid,
the  auxiliary function can be rewritten as
\begin{eqnarray}
        \tilde S(\Delta k, K)
& = &
        \tilde H(\Delta E) \int d^3  r \,\,
         \exp\left( - i \bbox{k} \bbox{\cdot} \bbox{r} \right) \,
         f(t_0,\bbox{r}; K),
\label{eq_AuxS}
\end{eqnarray}
where $\tilde H(\Delta E)$ stands for
the Fourier-transformed freeze-out time distribution.
Within this approximation the freeze-out time distribution
determines the energy-difference dependent part
of the correlation function.

%.................................................................
\subsection{Extension to  Long Emission Times}
\label{sec_LoEmTi}
%.................................................................
For low and intermediate energy nuclear reactions
the duration time for the particle emission can be quite long.
Evaporative models (with long duration times)
have been fairly successful in describing certain aspects
of such reactions \cite{FriedLynch,PrattTsang}.
The model presented above can be extended
to also include long emission times
by introducing cooling of the source
and by modifying the time distribution.

The  effect of possible cooling of the source
can be caused, e.g., by the expansion
and by the evaporation of the particles from the source.
Although cooling can be consistently incorporated
in the model of section \ref{sec_modTGrad}
(based on hydrodynamical results)
it is very difficult to estimate analytically
the effects for a long-lived, slowly cooling source,
when particles are emitted during a large time interval.
We therefore instead choose to utilize
a phenomenological cooling profile,
i.e.\ we assume that the decrease of the local temperature
is given by
\begin{equation}
   T(\bbox{r},t)
=
   T(\bbox{r}) \left( \frac{ \tau }{ t } \right)^\alpha,
\label{eq_Tcool}
\end{equation}
where $T(\bbox{r})$ is given by Eq.\ (\ref{e:tprof})
and we use values
$\tau=100$ fm/$c$ and $\alpha = 1/3$,
motivated by a profile
for isentropic three-dimensional expansion
of and ideal gas. 
Note, that this form may not be a good approximation
for the initial stage of the reaction 
since such a form cannot describe
the initial rise of the temperature. 
However, we numerically found,
that a long particle evaporation time 
is needed to get a simultaneous description 
of particle spectra and correlations 
in intermediate energy heavy ion collisions.
Thus, most of the particles are emitted much later 
than the initial stage of the reaction 
and the approximate profile given by the above equation 
may be suitable to describe the fall of the temperature
during most of the particle emission. 
A more detailed temperature profile
could be obtained by a fit 
to the local temperature distribution 
of a microscopical simulation 
including collective effects (re-scattering)
and particle evaporation.

The Gaussian time distribution in Eq.\ (\ref{eq_H(t)_Gauss})
does not give a good description 
for small times, $t \approx 0$,
when $\Delta t$ is large, and thus has to be modified.
Closely related to the Gaussian approximation
is a distorted Gaussian distribution,
which vanishes for very small times.
We assume  the form
\begin{equation}
  H_{\Gamma}(t) = 2 \, t \, d \, \exp ( - \, t^2 \, d ) \theta(t)\ ,
\label{eq_H_G(t)}
\end{equation}
which has the mean emission time
\begin{equation}
  \langle t \rangle  = \frac{1}{2} \sqrt{ \frac{\pi}{d} }
\end{equation}
with the variance
\begin{equation}
  \sigma^2 (t) 
= 
  \langle t^2 \rangle - \langle t \rangle ^2 
= 
  \frac{1}{d} \left( 1 - \frac{\pi}{4} \right)\ .
\end{equation}
This form is motivated by the fact
that there must be an initial rise
in the particle emission
followed by a long exponential tail.
Such a distribution is for example
the gamma distribution in $t^2$,
which contains two parameters.
However, for very broad time distributions,
the mean and the variance are not very sensitive
to independent tuning of these two parameters.
Thus considering the current precision of the data
(that is discussed in section \ref{sec_ppAndnn}),
we have fixed the small $t$ behavior
of the $H_{\Gamma}(t)$ curve to a linear rise,
decreasing the number of free parameters by one
and simplifying the numerical integrations at the same time.
Note that the analytic expressions above
for $\langle t \rangle$ and $\sigma^2 (t)$
are valid only when cooling is excluded.
When cooling is included the mean emission time
and the variance will be somewhat modified
(see also the results and discussion
 in section \ref{sec_ppAndnn}).

The source function $S(x,p)$
with the time distribution, $H_{\Gamma}(t)$
in Eq.\ (\ref{eq_H_G(t)}),
describes an expanding source with long duration
of the particle emission,
including cooling of the source, Eq.\ (\ref{eq_Tcool}).
It thus describes the gross features of particle evaporation,
which is believed to take place at low collision energies.
However, as an evaporative model,
some features are only approximately treated.
For example, quantities like flow velocity ($\bbox{u}$)
and geometrical source size ($R_G$)
are taken time independent,
while in a more rigorous treatment
also such quantities would vary with the time.
Examples of more refined evaporative models
can be found in Refs.\ \cite{FriedLynch,Fried2}.
Here, however, we want to utilize a rather simple model,
containing few parameters,
applicable in a wide energy range
(at high energies particle evaporation is negligible).
We want a model that can describe nucleon emission
as well as pion emission,
and we want the model to be simple enough
so that certain analytic results can be derived.
Thus when the model is applied near its low energy limit
(as in section \ref{sec_ppAndnn}),
the extracted quantities like flow velocity and source radius,
will reflect time averages.

% ----------------------------------------------------------------
\section{Analytic approximations}
\label{sec_anaAppr}
% ----------------------------------------------------------------
The integral for the auxiliary function
$\tilde S(\Delta k, K)$
in equation (\ref{eq_AuxS}),
can be evaluated with the help
of the saddle-point method,
which is described in details
in Refs.\ \cite{uli,hhm:te,akkelin}.
At the saddle-point, the partial derivatives
of the emission function
with regard to either $r_x, r_y$ or $ r_z$
vanish simultaneously.

The saddle point, $\bbox{r}_s$ is found by assuming that
$ \mid \! \bbox{r}_s \! \mid \! / t_0 \ll 1$.
In this case one can expand the saddle point equations
around $\bbox{r} = 0$ and one can solve
the saddle point equations in a linearized problem.
Note that for $a = 0$ the method yields exact results
because the non-linearity of the saddle-point equations
is related to the non-vanishing values of $a$.
Since we utilize an expansion for
$\mid \! \bbox{r} \! \mid \! / t_0 \ll 1$,
all shapes of the temperature profile which lead
to the same second order expansion for small distances
as Eq.\ (\ref{e:tprof})
lead to the same results.

With the help of the above approximations,
we may rewrite the emission function as
\begin{eqnarray}
     S(x,p)
& = &
     c_g \, \exp\left(
      - \frac{ ( \bbox{p} - m \bbox{u}(\bbox{r}_s(\bbox{p}\,)) )^2 }
             {  2 m T(\bbox{r}_s(\bbox{p}\,)) }
      - \frac{ \bbox{r}_s(\bbox{p}\,)^2 }{ 2 R_G^2 }
      - \frac{ (\bbox{r} - \bbox{r}_s(\bbox{p}\,))^2 }
             { 2 R_*^2 } \right) \, H(t)\ ,
\end{eqnarray}
with
\begin{eqnarray}
       c_g
& = &
       \frac{ g }{ (2 \pi)^3 }   
       \exp\left(  \frac{ \mu_0 }{ T_0} \right)\ .
\end{eqnarray}
The mean emission point coincides with the saddle-point
                $\bbox{r}_s(\bbox{p}\,)$,
being
\begin{eqnarray}
        \bbox{r}_s (\bbox{p}\,)
& = &
        b \, t_0 \, \frac{ \bbox{p} }
                         { a^2 \,  E_{k} (\bbox{p}) + b^2 \, m +
                           t_0^2 \, T_0 / R_G^2  }  ,
\end{eqnarray}
where the kinetic energy is denoted by
$E_k(\bbox{p}) = \bbox{p}^{\, 2} / (2\, m)$.
The following result is obtained for $R_*$:
\begin{eqnarray}
        \frac{ 1 }{ R_*^2(\bbox{p}) }
& = &
        \frac{ 1 }{ R_G^2 } +  \frac{ 1 }{ R_T^2(\bbox{p}) },
\label{e:rsp0}
\\
        {R_T^2(\bbox{p})}
& = &
        t_0^2 \, \frac{ T_0 }{ a^2 E_k(\bbox{p}) + b^2 m }\ .
\label{e:rsp}
\end{eqnarray}
Within the above approximations,
the auxiliary function contains a momentum-dependent factor
which shall enter the momentum distribution only
and another factor which is a Fourier-transform
of a Gaussian and so easily integrable.
The approximations are self-consistent
if the condition
$\mid \! \bbox{r}_s \! \mid \! /t_0 \ll 1$
is satisfied.
We return to this point later.

The analytic results for the momentum distribution
and the quantum statistical correlation function
are given as
\begin{eqnarray}
    N_1(\bbox{p})
& = &
    c_g \,  \left( 2 \pi R_*^2(\bbox{p}\,)\right)^{(3/2)} \,
    \exp\left(
      - \frac{( \bbox{p} - m \bbox{u} (\bbox{r}_s(\bbox{p}\,)) )^2}
             {2 m T(\bbox{r}_s(\bbox{p}\,))}
      - \frac{ \bbox{r}_s(\bbox{p}\,)^2 }{ 2 R_G^2 } \right)\ ,
\label{e:imd}
\\
        C(K,  \Delta k\,)
& = &
        1 \pm \exp( - R_*^2( \bbox{K})
        \bbox{\Delta k}^2 - \Delta t^2 \Delta E^2)\ .
\end{eqnarray}
The effects of final state Coulomb and Yukawa interactions
on the two-particle relative wave-functions are neglected
(in section \ref{sec_ppAndnn} final state interactions
 are taken into account).

These expressions are generalizations
of the momentum distribution and correlation function
obtained in Ref.\ \cite{nr}
and reduce to those for $a=0$ and $b = 1$.
The static fire-ball corresponds to the case $a = b = 0$.
A new feature for $a \neq 0$
is that the radius parameter of the QSCF
becomes a decreasing function of the momentum
and also the effective temperature
becomes momentum dependent.

\begin{figure}
\centerline{\psfig{file=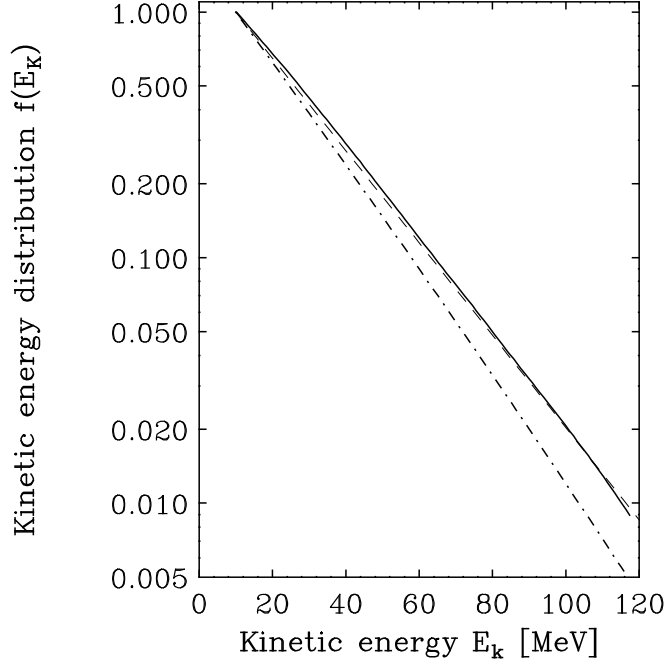,height=9.0cm}}
\caption{ Comparison of the numerically integrated
          energy distribution of protons (solid line)
          with a Boltzmann distribution of
          $T_* = 23.1$  MeV (dashed line)
          and with analytical results
          for the linearized saddle-point calculation,
          Eq.\ (\protect\ref{e:imd})
          as indicated by the dash-dotted line,
          for the parameter values of
          $ m = 938.3$ MeV$/c^2$,  $ R_G = 7.6 $ fm,
          $T_0 = 10.0$ MeV,        $t_0 = 45.0$ fm/c,
          $a = 1.0$,        and    $ b = 0.7 $.
          The energy distributions,
          $f(E_k) \propto N_1(\protect\bbox{p}) /\protect\bbox{p}^2$,
          are re-scaled so that $f(E_{k,0} ) = 1$
          for $E_{k,0} = 10 $ MeV.
          Note the characteristic bending down
          of both the solid and the dash-dotted line
          as compared to the Boltzmann distribution.
        }
\label{fig_f(E)}
\end{figure}

Similarly to the $a = 0$ and $b = 1$ case \cite{nr},
the effective temperature $T_*$ shall be determined
by the maximum of the local temperature, $T_0$,
and the geometrical temperature defined as
\begin{eqnarray}
        T_G & = & T_0 \frac{ R_G^2 }{ R_T^2 }\ .
\end{eqnarray}
	The relationship is given by
\begin{eqnarray}
        \frac{ 1 }{ T_* }
& = &
        \frac{ f }{ T_0 + T_G } + \frac{ 1 - f }{ T_0 }\ ,
\\
	f
& = &
        \frac{ b^2 }{ a^2 + b^2 }
\end{eqnarray}
which is analogous to the case
obtained for the slope of the momentum distribution
at high energies \cite{3d,3d:2}.
In the case of no temperature gradient,
$T_*$ will grow linearly with the particle mass.
In case of no flow, $T_*$ is independent of mass.
For a given mass, the effective temperature $T_*$
shall be constant only in a limited $p$ interval,
which is given by
$ p^{2} \ll m^2 b^2 / a^2$.
Since the present investigation
is limited to the non-relativistic $ p \ll m $ region,
it follows that the effective temperature
is not noticeably momentum-dependent
for small temperature gradient satisfying
$a^2 \leq b^2$.
As soon as the temperature gradient
increases above the $ a^2 \leq b^2$ region,
a characteristic high momentum suppression may appear
in the tail of the distribution,
see Fig.\ \ref{fig_f(E)} for example.

When fitting the model to data,
we  concentrate on the $a^2 \ge 0$ case,
since this is the case which may result
in a suppression at high momentum --
a phenomenon observed in the momentum distribution of protons 
in intermediate energy heavy ion reactions \cite{chitwood,bo}.

A short discussion on the validity
of the saddle-point approximation
is presented in appendix \ref{sec_appA},
together with suggested improvements of the approximation
and alternative approximation schemes.

% ----------------------------------------------------------------
\section{Correlations without final state interactions}
\label{sec_C(q)wNoFSI}
% ----------------------------------------------------------------

In this section the analytic results will be applied
to momentum distributions and correlation functions
where final state interactions can be neglected.
Applications could for example be
pion distributions and correlations at BEVALAC energies,
though the discussion in this section is quite general
and could be applied also to other situations
where final state interactions can be ignored.
The qualitative discussion in this section
also serves as a preparation to section \ref{sec_ppAndnn},
where final state interaction is introduced as well.
The presentation will follow the lines of Ref.\  \cite{nr},
since the results are very similar.

The relative momentum, $\bbox{\Delta k}$,
which appears in the  correlation function,
is invariant under Galilei-transformations,
but the energy difference is not invariant
even under the non-relativistic Galilei transformations.
This can be re-formulated
so that the specific directional dependence
becomes more transparent.
The energy difference is
\begin{eqnarray}
       \Delta E
=
       \frac{ \bbox{p}_1^{\, 2} - \bbox{p}^{\, 2}_2 }{ 2 m }
& = &
       \left( \frac{ \bbox{p}_1 + \bbox{p}_2 }{ 2 } \right) \cdot
       \frac{  \bbox{\Delta k} }{ m }
=
       \frac{ \bbox{K} \cdot \bbox{\Delta k} }{ m }
=
       \bbox{v}_{\bbox{K}} \cdot \bbox{\Delta k},
\end{eqnarray}
where we have introduced the mean velocity of the pair,
$\bbox{v}_{\bbox{K}} = \bbox{K}/m$.

Let us define the {\it out} direction
to be parallel to the mean velocity of the pair,
$\bbox{v}_{\bbox{K}}$,
and the {\it perp} index for
the remaining two principal directions,
both being perpendicular to the {\it out} direction.
This naming convention corresponds to the one
used in high energy heavy ion collisions \cite{bertsch,lutp}.

By this definition, the principal directions, which will be
utilized to evaluate the correlation function,
are assigned to the given mean momentum
of a given particle pair and {\it not} that of the fire-ball:
by changing the mean momentum we  change the {\it out} and the
{\it perp} directions as well.
In any given frame, the relative momentum can be decomposed as
\begin{eqnarray}
        \bbox{\Delta k}  &  = &  \bbox{Q}_{out} + \bbox{Q}_{perp}
\end{eqnarray}
and the correlation function can be rewritten as
\begin{eqnarray}
    C(\bbox{Q}_{out},\bbox{Q}_{perp})
& = &
    1 \pm  \exp\left( - R_{perp}^2(\bbox{K}) \bbox{Q}_{perp}^2
                   - R_{out}^2(\bbox{K}) \bbox{Q}_{out}^2 \right),
\label{e:cq}
\\
      R_{perp}^2(\bbox{K})
& = &
      R_*^2(\bbox{K})  \,
  =
      \, \frac{ R_T^2(\bbox{K}) R_G^2 }{ R_T^2(\bbox{K}) + R_G^2 },
\label{e:rp}
\\
        R_{out}^2(\bbox{K})
& = &
        R_*^2(\bbox{K}) + \bbox{v}_{\bbox{K}}^{\, 2} \Delta t^2\ .
\label{e:ro}
\end{eqnarray}
Thus the perpendicular components
shall measure the effective radius $R_*(\bbox{K})$.
What is the interpretation of this quantity?

\begin{figure}
\centerline{\psfig{file=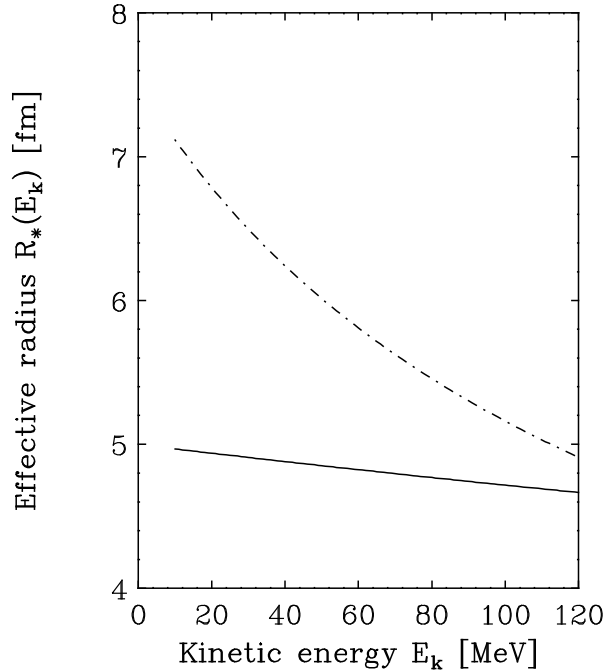,height=9.0cm}}
\caption{ Kinetic energy dependence
          of the effective radius parameter $R_*(E_k)$,
          Eq.\ (\protect\ref{e:rsp0}-\protect\ref{e:rsp}).
          Solid line indicates $R_*(E_k)$
          for the parameter values of Fig.\ \protect\ref{fig_f(E)},
          dash-dotted line stands for
          $ a = 2.0$, $ b = 0.1$,
          other parameters being the same as for the solid line.
        }
\label{fig_R*}
\end{figure}

There are two length-scales in the problem:
the geometrical length-scale $R_G$
and the thermal one $R_T$,
the latter being generated by the flow gradient $ b/t_0$,
the central temperature $T_0$
and the temperature gradient $a/t_0$.
{}From the previous equation one can see
that the effective $R_*(\bbox{K})$,
measured by the {\it perp} component
of the correlation function,
is dominated by the smaller of the two.
In other words,
for large and relatively cold
three-dimensionally expanding systems,
not the whole source
can be seen by quantum statistical correlations,
but only a part of the whole system,
which is characterized
by a thermal length-scale $R_T(\bbox{K})$.
Note also that, unlike in the case discussed in Ref.\ \cite{nr},
the radius parameter $R_*(\bbox{K})$ shall not be a constant
but it will depend on the momentum of the particles.
See Fig.\ \ref{fig_R*} for illustration.

{}From Eq.\ (\ref{e:ro}) it follows
that the out component in general
shall also be sensitive to the duration
of the freeze-out time distribution,
since it contains a term
$\bbox{v}_{\bbox{K}}^{\, 2} \Delta t^2 $.
This term vanishes in the
{center of mass frame (c.m.f.)}
of the particle pair, since
$\Delta E = \mid \! \bbox{v}_{\bbox{K}} \! \mid = 0$
in c.m.f.
In this specific system,
\begin{eqnarray}
        Q_I^2
	= \bbox{Q}_{perp}^2 + \bbox{Q}_{out}^2 - \Delta E^2 =
        \bbox{Q}_{perp}^2 + \bbox{Q}_{out}^2  \qquad
        \mbox{\rm in~c.m.f.~of~the~pair}
\end{eqnarray}
is the invariant momentum difference.
The correlation function in the considered case
becomes symmetric, when evaluated in the c.m. of the pair:
\begin{eqnarray}
        C(Q_I)
& = &
        1 \pm \exp( - R_*^2(\bbox{K}) Q_I^2)  \qquad
        \mbox{\rm in~c.m.f.~of~the~pair}.
\end{eqnarray}

It is interesting to investigate another limiting case,
$R_G \ll R_T(\bbox{K})$.
In this case we re-obtain the standard results:
\begin{eqnarray}
        R_{perp}^2
=
        R_G^2\ ,
\qquad
        R_{out}^2
=
        R_G^2 + \bbox{v}_{\bbox{K}}^2 \Delta t^2\ ,
\qquad
        T_*
=
        T\ ,
\end{eqnarray}
i.e., if the thermal length scale
is larger than the geometrical size in all three directions,
the correlation measurement
determines the geometrical sizes properly,
and the momentum distribution
will be determined by the freeze-out temperature
being just a thermal distribution for a static source.
In this limiting case, the effective radius parameter
as well as the effective temperature
become independent of the momentum of the emitted particles.

The non-relativistic model presented in this work
has six input parameters:
$R_G$, $t_0$, $\Delta t$,  $T$, $ a $ and $b$,
which determine two measurable {\em functions}.
This, in turn, implies that all these parameters
can in principle be determined from the detailed analysis
of the momentum dependence
of the two-particle correlation functions
and from the deviation
of the one-particle invariant momentum distributions
from the purely thermal behavior.

For the special case when the measured QSCF-s
are found to be momentum-independent
and the IMD is well described
by a Boltzmann momentum distribution,
the $a = 0$ case is concluded.
In this case, we may measure the parameters
$\Delta t$, $R_*$ and $T_*$
which are constants.
The measurable parameters
$R_*$ and $T_*$
are determined by four model parameters,
$R_G$, $T_0$, $t_0 $ and $b$.
Thus the model parameters
cannot be determined uniquely
from the IMD and QSCF observables
in the $a=0 $ special case.
Inequalities can be obtained,
along the lines of Ref.\ \cite{nr},
to restrict the values of the model parameters
in this special case.
For example, we have
$T_0 \le T_*$ and $R_G \ge R_*$.
Note that a similar degeneracy
of $T_*$ and $R_*$
has been found in Ref.\ \cite{nr}
for the $a=0$ and $b=1$ special case,
where it was observed that the
the same spectra and correlations can be obtained
for different values of the three input parameters.

The general results for the correlation function
given in Eqs.\ (\ref{e:cq}-\ref{e:ro})
indicate structural similarity
with the Bose-Einstein correlation function parameters
for a class of models
which includes relativistic longitudinal flows,
non-relativistic transverse flows
and a transverse temperature profile \cite{3d,3d:2}.
The radius parameters of the correlation function
are momentum dependent
both for the non-relativistic model presented here
and for the model-class discussed in Refs.\ \cite{3d,3d:2}.
The structural similarity implies
that the effective duration parameter $\Delta t_*^2$ ,
defined as the coefficient of $\bbox{v}_{\bbox{K}}^{\, 2}$
in Eq.\ (\ref{e:ro}),
shall become momentum dependent:
$\Delta t_* = \Delta t_*(\bbox{K})$
if the cooling effects are switched on.
  This new, momentum-dependent
duration parameter $\Delta t_*(\bbox{K})$ shall replace the
 momentum-independent $\Delta t$ duration parameter
 in the correlation function if the temporal changes of the
 temperature are significant.

% ---------------------------------------------------------------
\section{Application to Neutron and Proton Interferometry}
\label{sec_ppAndnn}
% ---------------------------------------------------------------

In section \ref{sec_C(q)wNoFSI} it was discussed
that the effective radius parameter $R_*(\bbox{K})$
is not constant but depends on the momentum of the particles.
Such an effect has indeed been seen
in the measured proton-proton correlation functions
in the $^{27}{\rm Al}\,( ^{14}{\rm N},\,pp)$ reactions
at $E = 75$ MeV/nucleon \cite{bauer,66}:
the larger the momentum of the protons
the smaller the effective source size \cite{bauer}.
This feature is in qualitative agreement
with the analytic results given in section \ref{sec_anaAppr},
since the thermal radius $R_T(\bbox{K})$
is a decreasing function of the absolute value
of the mean momentum, $|\bbox{K}|$.
The effective source size is dominated
by the smaller of the thermal and geometrical radius,
thus it is also a decreasing function
of the mean momentum of the particle pair.

The results presented in the previous sections
contain the essential ingredients
which are needed to obtain a momentum-dependent radius parameter
in the non-relativistic domain.
However, to  be utilized for nucleons
emitted in intermediate energy 
heavy ion reactions \cite{boal,bauer},
also two other important effects need to be considered,
namely cooling of the source
(see section \ref{sec_LoEmTi})
and final state interactions.
Final state interactions can be included for $pp$ pairs
utilizing the Coulomb + strong interactions
and Fermi-Dirac statistics,
the strong interactions and Fermi-Dirac statistics for $nn$ pairs
and the strong interactions only for the $np$ pairs,
utilizing the Wigner-function formalism developed
by S. Pratt and collaborators
(see e.g. Refs.\ \cite{bauer,pratt_csorgo}
for the description of the method).
Essentially, this calculation includes the evaluation
of the two-particle relative wave-function
with the above Coulomb and/or strong interaction
using the appropriate quantum statistical
{(anti)symmetrization}
and averaging the result over the particle distribution
calculated from the model in section \ref{sec_modTGrad}.
A Reid soft core potential is used to take into account
the strong final state interactions.
The essential consequences of the final-state interactions
are that they completely modify the short-range parts
of the two-particle correlations,
and they create a peak at low relative momentum
in the correlation functions.
In the case of proton pairs,
this peak is suppressed by the Coulomb repulsion
which creates a hole centered at zero relative momentum
in the $pp$ correlation function.

Light particle interferometry
at
{intermediate energy}
heavy-ion collisions
has been extensively studied,
both experimentally and theoretically.
Many attempts,
using different models have been put forward,
for example:
simple Gaussian source parameterizations,
evaporation,
and transport models such as QMD and BUU
(see Refs.\ \cite{bauer,ardouin} and references therein).
These models,
containing different physical ingredients and information,
have to different degrees been successful
in describing the data.

One of the purposes of this paper
is to test if the model,
presented in section \ref{sec_model},
is applicable to $nn$ and $pp$ interferometry
at intermediate energies.
The model contains a rather small set of parameters,
and by making simultaneous fits
to $n$ and $p$ single energy spectra,
and $nn$ and $pp$ correlation functions,
rather hard constraints are put on the parameter set.
Thus a qualitative and quantitative understanding
of certain aspects of the source can be extracted.

The evolution of the particle emission
in a heavy-ion collision at intermediate energies
may roughly be described as:
 production of pre-equilibrium particles;
 expansion and possible freeze-out of a compound source;
 possible evaporation from an excited residue of the source.
Note though, that this separation is not very distinct
and there is an overlap between the different stages.
The importance of the various stages above
also depends on the beam energy
and the impact parameter of the collision.
The model presented in this paper describes well
the second stage above
and, for long emission times, also part of the third stage.
If these stages give the main contribution
to the emission of nucleons,
a satisfactory description of experimental data
can be obtained.

\begin{table}[h]
\begin{tabular}{|l|c|c|c|c|c|}
           &   $R_G$ (fm)                 &   $T_0$ (MeV)
           &   $a/t_0$ (fm/$c)^{-1}$      &   $b/t_0$ (fm/$c)^{-1}$
           &   $d$ (fm/$c)^{-2}$
\\
\hline
\hline
Neutrons   &   4.0                        &   3.0
           &   0.0                        &   0.018
           &   $5.0 \cdot 10^{-5}$
\\
\hline
Protons    &   4.0                        &   5.0
           &   0.14                       &   0.036
           &   $5.0 \cdot 10^{-5}$
\\
\end{tabular}
\caption{ Parameter values used for calculating
          $n$ and $p$ spectra and correlation functions }
\label{tab_1}
\end{table}
We have applied our model to the reaction
$^{40}$Ar + $^{197}$Au at 30 MeV/nucleon,
to compare with experimental single spectra
and correlation functions
from the $np$ correlation experiments
described in Refs.\ \cite{bo,Cronq,Ghetti}.
With the parameter set presented in table \ref{tab_1}
we have obtained a simultaneous fit
to $n$ and $p$ single spectra,
as well as $nn$ and $pp$ correlation functions.
Thus we have used the same parameters
for neutrons as for protons,
except for the parameters $T_0$, $a$ and $b$.
As discussed in sections \ref{sec_introd} and \ref{sec_model}
the proton energy spectrum deviates
from the thermal spectrum \cite{chitwood,bo},
although the neutron spectrum is well approximated
by a Boltzmann distribution.
The Coulomb interaction makes a difference
between protons and neutrons,
and this behavior is effectively obtained
within our framework for protons by allowing
a temperature gradient inside the source
and a different flow parameter.
Protons emitted from the middle of the source
roll down from a higher Coulomb-potential
than those from the surface,
thus protons from the middle are emitted
with a higher kinetic energy
than those emitted from the surface.
This qualitative feature is similar
to a system which is hotter in the middle
than at the surface.
Incorporating a temperature gradient inside the source
does effectively describe such an effect.

To keep the description as simple as possible
in this first attempt to test the applicability
of the model at intermediate energies,
we have ignored impact parameter averaging
and Coulomb interaction with the source.
An averaging over different impact parameters
could for example be done
by allowing the radius parameter, $R_G$,
to vary with the geometrical overlap
of the projectile and target.
However, such a prescription
also introduces an additional uncertainty,
and has therefore been ignored
to keep the description simple.
Thus the extracted source size
will reflect a ``mean'' source size
(assumed to be averaged 
 over the different impact parameters).

When performing the calculations
we have taken into account
the experimental energy thresholds
and the acceptance region
of the experimental set up.

\begin{figure}[h]
\centerline{\psfig{file=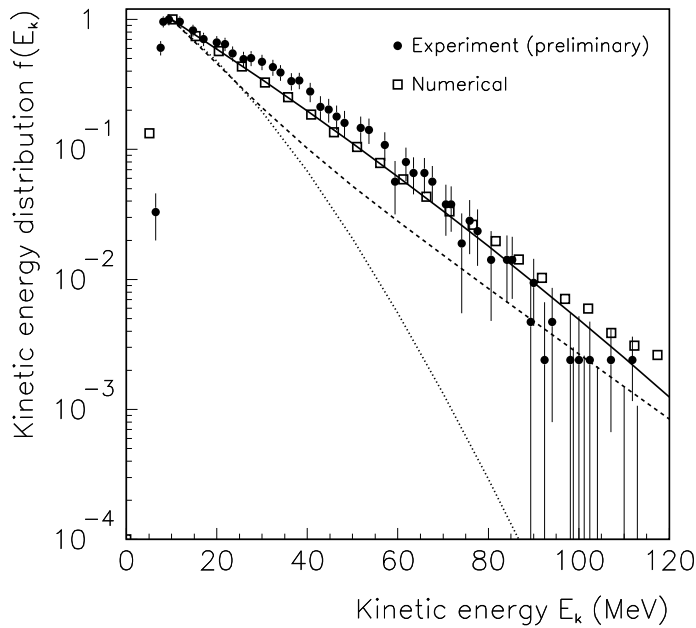,height=9.0cm}}
\caption{ Experimental and numerically calculated
          proton energy spectrum.
          The filled circles represent
          preliminary inclusive 
          proton emission data \protect\cite{bo}
          from the $np$ correlation experiment
          described in Ref.\ \protect\cite{Cronq},
          the open squares the full numerical simulation,
          while analytic approximations are represented by
          a solid curve (Eq.\ (\protect\ref{eq_Taylor})),
          dashed curve (Eq.\ (\protect\ref{eq_Saddle2}))
          and a dotted curve (Eq.\ (\protect\ref{e:imd})), 
          respectively.
          The experimental data are preliminary, 
          without fully estimated errors, 
          why we have estimated the errors 
          from the fluctuations of the data points.
          The parameter set in table \protect\ref{tab_1} 
          was employed for the numerical calculations.
          The energy distributions,
          $f(E_k)$ are re-scaled so that $f(E_{k,0} ) = 1$
          for $E_{k,0} = 10 $ MeV.
        }
\label{fig_fp(Ek)}
\end{figure}
Experimental inclusive proton emission data \cite{bo}
(from the $np$ correlation experiment
described in Ref.\ \cite{Cronq})
are presented in Fig.\ \ref{fig_fp(Ek)} (filled circles)
together with our fit (open squares).
In addition we show the analytic approximate expression
in Eqs.\ (\ref{e:imd}), (\ref{eq_Saddle2}) and (\ref{eq_Taylor})
as a dotted, dashed and solid curve respectively.
The neutron spectrum, not shown, is purely thermal
with an effective temperature of 8.5 $\pm$ 1 MeV
and is also well reproduced by our calculations.

\begin{figure}[h]
\centerline{\psfig{file=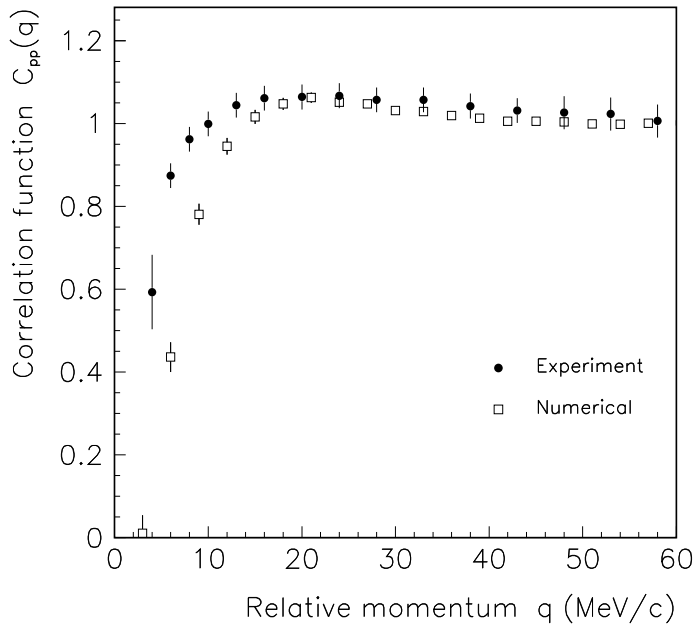,height=9.0cm}}
\caption{ Experimental and numerically calculated
          $pp$ correlation function.
          The filled circles represent data
          from the $np$ correlation experiment
          described in Ref.\ \protect\cite{Cronq},
          the open squares the full numerical simulation.
          The parameter set in table \protect\ref{tab_1} 
          was employed for the numerical calculations.
        }
\label{fig_Cpp(q)}
\end{figure}

Proton correlation functions are presented
in Fig.\ \ref{fig_Cpp(q)}
as a function of the relative momentum
$q=|\bbox{p}_1-\bbox{p}_1|/2$,
integrated over the total pair momentum
$(\bbox{p}_1 + \bbox{p}_2)$.
The experimental results are reproduced qualitatively,
in some regions of the momentum space even quantitatively.
Considering an experimental uncertainty in $q$
of $3.0 \leq \Delta q \leq  5.5$ MeV/$c$ \cite{Ghetti},
which is not taken into account in our calculations,
and the known difficulty
to describe both the single-particle spectrum
and the correlations for protons and neutrons
at the same time \cite{upe},
we think that our result is interesting.
The discrepancy for $q < $ 10-15 MeV/$c$
is partly due to the experimental uncertainty $\Delta q$.
However, a discrepancy  has also been seen
in other models (see e.g. Ref.\ \cite{ardouin})
and suggested explanations are the following:
 more than one source,
 Coulomb interaction with the source or
 pre-equilibrium emission.

\begin{figure}
\centerline{\psfig{file=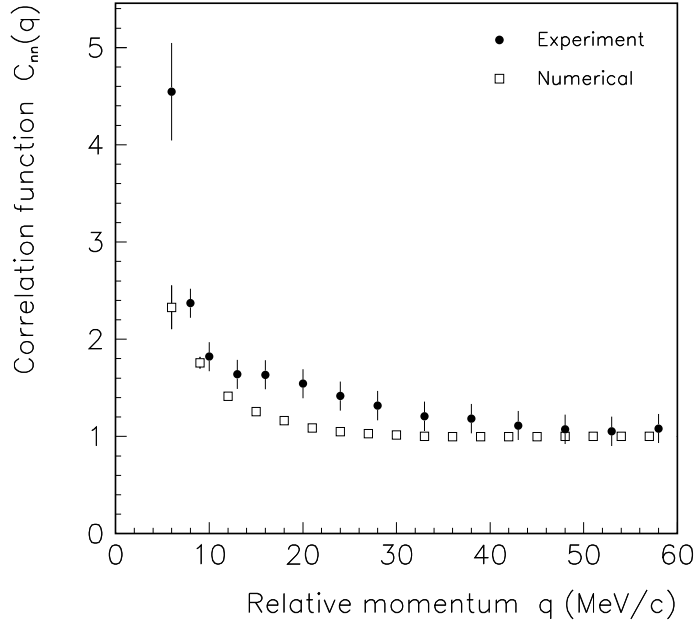,height=9.0cm}}
\caption{ Experimental and numerically calculated
          $nn$ correlation function.
          The filled circles represent data
          from the $np$ correlation experiment
          described in Ref.\ \protect\cite{Cronq},
          the open squares the full numerical simulation.
          The parameter set in table \protect\ref{tab_1} 
          was employed for the numerical calculations.
        }
\label{fig_Cnn(q)}
\end{figure}

The experimental $nn$ correlation function
contains some uncertainties,
namely cross talk between the detectors,
and experimental cuts at low relative momentum.
{}From simulations, the cross talk
(see Fig.\ 3e of Ref.\ \cite{Cronq} )
has been found to mainly contribute
to the ``bump'' around $15 < q < 40$ MeV/$c$.
Furthermore the sharp rise in $C(q)$
at $ q \approx 10 $ MeV/$c$ is based
on rather few events with neutron kinetic energy
smaller than about 10 MeV
(see Fig.\ 3d of Ref.\ \cite{Cronq})
close to detector energy threshold and acceptance limits.
The momentum uncertainty, $\Delta q$,
is smaller than 2 MeV/$c$.
Considering these uncertainties
the experimental $C(q)$ is reproduced
in an acceptable manner in Fig.\ \ref{fig_Cnn(q)}.

When performing the parameter fit
we have found that it is not possible
to simultaneously reproduce $n$ and $p$ energy spectra,
as well as $nn$ and $pp$ correlation functions,
unless the duration of the particle emission is taken large
(several hundreds of fm/$c$).
The long duration is found to be responsible 
for the low relative momentum behavior
of the two-particle correlations.
Thus we have used $H_{\Gamma}(t)$ in Eq.\ (\ref{eq_H_G(t)})
for the time distribution.
With the parameter set in table \ref{tab_1}
we obtain numerically the mean emission time $\langle t \rangle$
and duration $\sigma(t)$:
\[
    \langle t \rangle       \approx 520 \mbox{ fm}/c \qquad
    \sigma(t) \approx 320 \mbox{ fm}/c \ .
\]
(Note that $\langle t \rangle$ necessarily becomes large
because of the broad time distribution limited to $t>0$.)
This scenario is in agreement with the general view
of nuclear reactions in the energy domain
of a few tenths of MeV/nucleon,
namely a rather long-lived excited source
emitting particles.
Note though that a pure evaporative source
gives a worse description
of the experimental $nn$ and $pp$ correlation functions
(see Ref.\ \cite{Cronq})
and fails to reproduce the single particle energy distribution.
Thus the inclusion of the expansion seems to be mandatory,
although the extracted flow velocities are small.

The extracted geometrical source radius $R_G \approx 4$ fm
is quite reasonable considering
that a Gaussian parameterization is used
(instead of a Woods-Saxon distribution,
usually employed for the ground state density distribution).
Note also that the extracted source size
reflects an average over different times
and impact parameters.
The effective radius, $R_*$, seen in the correlation function
is smaller than $R_G$;
1.1 - 1.8 fm for protons (depending on momentum)
and 2.5 fm for neutrons.

{}From the experimental energy spectra
we know that the effective temperature for neutrons
should be 8.5 $\pm$ 1 MeV
and for protons in the range 15 - 25 MeV
depending on the energy.
The parameter set in table \ref{tab_1} gives
\[ T_{\rm eff} = T_* \approx 7.9 \mbox{ MeV} \]
for neutrons,
and for protons,
using the Taylor approximation in Eq.\ (\ref{eq_Taylor}),
we obtain
\[
        T_{\rm eff} \approx 20.1 \mbox{ MeV},
\quad
        E_k = 10 \mbox{ MeV};
\qquad
        T_{\rm eff} \approx 17.4 \mbox{ MeV},
\quad
        E_k = 100 \mbox{ MeV}\ .
\]
This is obtained with an input temperature
of $T_0$ = 3 MeV for neutrons,
and for protons with a temperature profile
that decreases from 5 MeV at the center of the source
to 4.3 MeV at $r=R_G$ and 3.1 MeV at $r=2R_G$.

The parameter set also gives the flow velocities
\[ u(n) \approx 0.018 r c; \qquad u(p) \approx 0.036 r c\ . \]

The results presented in Figs.\
\ref{fig_fp(Ek)} -- \ref{fig_Cnn(q)}
are obtained with cooling included
according to Eq.\ \ref{eq_Tcool}.
We have numerically found
that the results change only to a minor extent
if the cooling is excluded.
Thus the different approximations in Eqs.\
(\ref{e:imd}), (\ref{eq_Saddle2}) and (\ref{eq_Taylor})
are useful also when cooling is included.

A little note should be made on the parameter fit in our model.
It is not possible to determine the parameters uniquely
from the experimental data,
though they are strongly constrained.
Further constraints could be imposed
by also investigating the dependence
of the correlation function
on the total pair momentum,
though the current data set does not allow
such an investigation.

% ---------------------------------------------------------------
\section{Summary}
\label{sec_Sum}
% ---------------------------------------------------------------

In summary, we have calculated
the invariant momentum distribution
and the quantum statistical correlation function
in a non-relativistic model
for spherically symmetric non-relativistic expansion
with possible temperature gradient,
phenomenologically including cooling
and particle evaporation.
We have thus presented a generalization
of the previous work \cite{nr}
by introducing new parameters.
These parameters control
 the strength of the flow,
 the temperature gradient,
 the strength of cooling
 and the duration of particle emission.

We have observed structural similarity
of this non-relativistic model
to the relativistic expansion
described in Refs.\ \cite{3d,3d:2},
namely the effective radius parameters
of the two-particle correlation functions
and the effective slope parameters
of the single particle spectra
became momentum dependent
due to the interplay of the local thermal scales
and the geometrical scales.
On this level, the model presented here
is very similar to the relativistic one.

The main effects of the temperature gradient
are that it introduces
 {\it i)}
   a momentum-dependent effective temperature
   which is decreasing for increasing momentum,
   resulting in a suppression at high momentum
   as compared to the Boltzmann distribution; 
 {\it ii)}
   a momentum-dependent effective source size
   which decrease with increasing total momentum.
These qualitative features have been seen
in non-relativistic heavy-ion collisions.

The model presented here,
including final state interactions,
has been applied to measured correlation functions
and preliminary neutron and proton energy spectra
in the reaction
$^{40}$Ar + $^{197}$Au at 30 MeV/nucleon \cite{Cronq}.
Agreement with the experimental data is obtained
only if the duration time
of the particle emission is large.
The obtained parameter set reflects
a moderately large system
(Gaussian radius parameter $R_G$ = 4.0 fm)
at a moderate temperature
($T_0(n)$ = 3 MeV and $T_0(p)$ = 5 MeV)
and small flow.
Note, however,
that the agreement between the model and the data
was obtained only if the flow effect is included,
i.e.\ within this phenomenological picture
the inclusion of some flow is needed.

% ---------------------------------------------------------------
\acknowledgments
% ---------------------------------------------------------------
J.H.\ would like to thank
 B.\ Jakobsson, R.\ Ghetti and the CHIC collaboration
 for helpful discussions,
Cs.T.\ would like to thank
 D. Ardouin, B.\ Jakobsson and G.\ Gustafson
 for helpful discussions and
 M.\ Gyulassy and X.-N.\ Wang
 for kind hospitality.
S.\ Pratt is acknowledged for his comments on the manuscript
and for making his interferometry code available to us.
This work was supported
by the Training and Mobility through Research (TMR) programme
 of the European Community under contract ERBFMBICT950086
 and by the Swedish Natural Science Research Council,
by the Human Capital and Mobility (COST) programme of the
 EEC under grant No. CIPA - CT - 92 - 0418 (DG 12 HSMU) and
by the Hungarian NSF
 under Grant  No. OTKA-F4019 and W/01015107,
by an Advanced Research Award
 of the Fulbright Foundation,
by the Director, Office of Energy Research,
 Office of High Energy and Nuclear Physics,
 Nuclear Physics Division of the U.S. Department of Energy
 (Contracts No.  DE-AC03-76SF00098 and  No.DE-FG02-93ER40764).

% ---------------------------------------------------------------
\appendix
\section{Details of analytic approximations}
\label{sec_appA}
% ---------------------------------------------------------------

We have examined the validity
of the saddle point approximation,
presented in section \ref{sec_anaAppr},
both analytically and by comparing
with a numerically generated particle distribution.
The saddle point approximation
normally yields a good approximation
for large values of the expansion parameter (here $a$).
However, for some parameter sets
the saddle point approximation
is not a good approximation for large $a$ values
(and large momenta $p$).
This is because for large $a$,
the emission function can
develop more than one maximum,
and the second order expansion around one of the maxima
yields a poor approximation.
The range of values of $a$ and $p$ for which
the saddle point approximation is valid,
depends on the other parameters used.

For the case of two well separated maxima
the saddle point approximation can be improved
by summing up the contribution from the two maxima.
Here we give the expression for the momentum distribution
obtained in this approximation
\begin{eqnarray}
      N_1(\bbox{p})
& \approx &
      c_g \, \left( 2 \pi R_*^2(\bbox{p}\,)\right)^{(3/2)} \,
      \left\{ \exp\left( - \frac{E_k}{T_{*0}}
                   \left[ \frac{ 1+a^2 E_k R_G^2/T_0t_0^2 }
                               { 1+a^2 E_k R_{*0}^2/T_0 t_0^2 }
      \right] \right) \right.
\nonumber \\
& & +
      \left.  \exp\left( - \frac{E_k}{T_{*0}}
                   \left[ \frac{ 1+a^2 E_k R_{T0}^2/T_0 t_0^2 }
                               { 1+a^2 E_k R_{*0}^2/T_0 t_0^2 }
      \right] \right) \right\}
\label{eq_Saddle2}
\end{eqnarray}

An approximate expression can also be obtained
for small values of the parameter $a$
by making a Taylor expansion
of the emission function around $a=0$.
Within this approximation we obtain
(identifying the integrated result
 as the first terms in an expansion
 of the exponential function):
\begin{eqnarray}
      N_1(\bbox{p})
& \approx &
      c_g \, \left( 2 \pi R_*^2(\bbox{p}\,)\right)^{(3/2)}
\nonumber \\
& \times &
      \exp\left( - \frac{E_k}{T_{*0}}
              - \frac{ 15 a^2 R_{*0}^4 }{ 4 t_0^2 R_{T0}^2 }
              - \frac{ a^2 E_k }{ T_0 t_0^2 } \left[
                  \frac{3R_{*0}^2}{2} -
                  \frac{ 5 R_{*0}^6 }{ R_{T0}^2 R_G^2 } \right]
              - \frac{ a^2 E_k^2 }{ T_0^2 t_0^2 }
                  \frac{ R_{*0}^8 }{ R_{T0}^2 R_G^4 }
          \right)
\label{eq_Taylor}
\end{eqnarray}
In the above expressions we have used the notation
\[ R_{T0} = R_T(\bbox{p}=0), \qquad
   R_{*0} = R_*(\bbox{p}=0)  \quad \mbox{ and } \quad
   T_{*0} = T_*(\bbox{p}=0) \ .
\]

Comparing these analytic approximations
with numerical calculated momentum distributions
we have found that for most parameter sets
a good agreement can be found
with either of the approximations.

% ---------------------------------------------------------------

% ---------------------------------------------------------------
%%%%%%%%%%%%%%%%%%%%%%%%%% Figures %%%%%%%%%%%%%%%%%%%%%%%%%%%%%%
% ---------------------------------------------------------------

% ---------------------------------------------------------------
%%%%%%%%%%%%%%%%%%%%%%%%%% Table %%%%%%%%%%%%%%%%%%%%%%%%%%%%%%%%
% ---------------------------------------------------------------

\end{document}